\documentclass[12pt,preprint]{aastex}

\usepackage{apjfonts}
\usepackage{emulateapj5}
\usepackage{amsmath}
\usepackage{psfig}
\usepackage{epsfig}
\usepackage{subfigure}
\usepackage{amssymb}

\newcommand{\ptl}{\partial}
\def\mp{m_{\rm p}}
\renewenvironment{figure}{\begin{figure*} }{\end{figure*}}

\renewcommand{\mp}{m_{\rm p}}
\newcommand{\blde}{{\bf e}}

\renewcommand{\bv}{{\bf v}}

\newcommand{\br}{{\bf r}}
\newcommand{\Fd}{{\bf F}_{\rm d}}
\newcommand{\grad}{{\mathbf \nabla}}
\newcommand{\rhos}{\rho_{\rm s}}
\newcommand{\rhop}{\rho_{\rm p}}
\newcommand{\vk}{v_{\rm K}}
\newcommand{\ok}{\Omega_{\rm K}}
\newcommand{\zp}{z_{\rm p}}
\renewcommand{\wp}{w_{\rm p}}
\newcommand{\dd}{{\rm d}}
\newcommand{\calH}{{\mathcal H}}
\newcommand{\etheta}{{ \hat{\blde}_{\theta}}}

\newcommand{\er}{{ \hat{\blde}_r}}
\newcommand{\ez}{{ \hat{\blde}_z}}

\newcommand{\vg}{v_{\rm g}}

\begin{document}

\title{Individual and collective behavior of dust particles in a protoplanetary nebula}
\author{P. Garaud, L. Barri\`ere-Fouchet, D. N. C. Lin}
\affil{Institute of Astronomy, Madingley Road, Cambridge, CB3 0HA, UK \\
Centre de Recherche Astronomique de Lyon, CNRS-UMR 5574, ENS Lyon, 46 allee d'Italie, F-69364 Lyon Cedex 07 France \\
UCO/Lick Observatory, University of California Santa Cruz, 1156 High Street, CA 95064 Santa Cruz, USA}

\maketitle

\begin{abstract}
We study the interaction between gas and dust particles in a
protoplanetary disk, comparing analytical and numerical results. We
first calculate analytically the trajectories of individual particles
undergoing gas drag in the disk, in the asymptotic cases of very 
small particles (Epstein regime) and very large particles (Stokes  
regime). Using a Boltzmann averaging method, we then infer their
collective behavior. We compare the results of this analytical
formulation against numerical computations of a large number of
particles. Using successive moments of the Boltzmann equation, we
derive the equivalent fluid equations for the average motion of the
particles; these are intrinsically different in the Epstein and Stokes 
regimes. We are
also able to study analytically the temporal evolution of a collection
of particles with a given initial size-distribution provided
collisions are ignored.
\end{abstract}

\section{Introduction}

In an attempt to account for the coplanar nature of the orbits of all
known solar-system planets, Laplace (1796) postulated that they were
formed in a common disk around the protosun.  Today, the detection of
protostellar disks around most young T-Tauri stars (Prosser {\it et
al.} 1994) is a strong evidence that the Laplace nebula hypothesis is
universally applicable. The recent discovery of planets around at
least 10\% of nearby solar-type stars (Marcy {\it et al.} 2000)
suggests that their formation may be a robust process.

Conventional cosmogonical scenarios are based on the assumption that
heavy elements in gas-phase condensed to form grains which then
coagulated into planetesimals and grew into protoplanetary cores which
can accrete, at least in some regions of the disk, massive gaseous
envelopes around themselves (Pollack {\it et al.} 1996).  The
coexistence of gas and solid ice has been detected in some
protostellar disks (Thi {\it et al.} 2002).  In fact, protostellar
disks are most conspicuous in their continuum radiation associated
with the re-processing of stellar light by the grains (Adams, Lada, \&
Shu 1987).  The apparent wavelength dependence in the thickness of
the disk dust layer has been interpreted as evidence of grain growth
(Throop {\it et al.} 2001, D'Alessio {\it et al.}, 2001, Clarke {\it et
al.}, 2003) and settling (Shuping {\it et al.}, 2003).

The $\mu$m-to-cm continuum radiation signatures of the dust are
observed to fade on the timescale of a few Myr (Beckwith, 1999, Haisch
{\it et al.} 2001), signaling the depletion of grains in this size
range.  This suggests that heavy elements initially contained in this
size range are either evaporated, ejected to large distance, accreted
onto the host stars, or have coagulated into larger particles.  The
first possibility is constrained by the concurrent decline in the
CO-gas (Zuckerman {\it et al.} 1995) whereas the last possibility is
directly relevant to the process of planet formation.

Theoretical analysis suggests a very strong constraint on the growth
of $\mu$m-size grains into km-size planetesimals.  Indeed, the orbital
evolution of the particles is determined by both the gravity of the
central star and the drag of the disk gas.  In the absence of
turbulence, the disk gas attains a dynamical equilibrium between
gravity, pressure, and centrifugal forces with zero velocity in both
radial and normal-to-the-disk directions and a slightly sub-Keplerian
velocity in the azimuthal direction.  Particles in the disk undergo
both sedimentation toward the mid-plane and inward drift in the radial
direction (Whipple 1972, Weidenschilling 1977).  In a minimum mass
nebula (Hayashi {\it et al.} 1985), the resulting orbital decay
timescale at 1AU (for instance) is smallest for m-size particles
(Adachi {\it et al.} 1976), and is then less than about $10^2$
yr. Unless the growth of planetesimals across this ``most vulnerable
size'' can occur faster than their orbital decay, there would be no
residual planetesimals left to provide the building blocks of planets.

One possible channel of rapid grain growth is through sedimentation
into a sufficiently thin, gravitationally unstable disk (Goldreich \&
Ward 1973).  The critical thickness for gravitational instability of
such disks is less than $\sim 10^{-5}$ of their radii and the
characteristic size of the resulting fragment is $\sim$ a few km.
However, even a modest amount of turbulence can provide adequate
stirring to prevent the sedimentation of grains into such a thin
unstable layer (Weidenschilling 1984, Supulver \& Lin 2000). Though
turbulence is likely to occur in a magnetized disk (Balbus \& Hawley,
1990) through magneto-rotational instability, this mechanism could
well fail in regions of the disk where the ionization fraction is too
small. In these regions only, the following alternative mechanism for
turbulence has been proposed.

In a laminar disk, the sedimentation of dust toward the disk's
mid-plane leads to a local concentration of massive particles; these
particles entrain the gas to a near-Keplerian velocity through drag,
thereby introducing a shear layer between the dust-dominated mid-plane
and the rest of the disk gas (Weidenschilling \& Cuzzi 1993). Such a
flow pattern in the disk has the potential to cause the onset of a
shearing instability (Sekiya 1998, Youdin \& Shu 2002).  However, the
stability analysis used by these authors for such flow is based on a
single-fluid approximation in which the dust particles are assumed to
be well-coupled to the gas.  Since the concentration of the dust
particles not only causes the shear but also a stabilizing density
stratification, the flow of dust and gas should be treated separately.
In a companion paper (Garaud {\it et al.} in preparation), we will
carry out a two-component stability analysis of the disk's dust layer.

Such a study is greatly simplified by the treatment of the particles
as a separate fluid rather than a collection of particles. It is with
this goal in mind that we now present a system of averaged equations
for the evolution of a collection of dust particles in the form of
moments of the Boltzmann equation.  This prescription could also in
principle be applied for the studies of dust particles' evolution due
to coagulation, sublimation, condensation (Supulver \& Lin 2000) and
under their interaction with embedded planets (Wyatt {\it et al.}
1999) and stellar radiation (Takeuchi \& Artymowicz 2001, Klahr \& Lin
2001, Takeuchi \& Lin 2002).

For the present calculation, we assume the particles are collisionless
and indestructible spheres in a gaseous laminar disk with no embedded
planets and negligible stellar radiation.  In this paper, we also
neglect the feedback of the particles' drag on the motion of the
gas. In \S2, we recall the general gas drag laws and their effects on
particle trajectories in a protoplanetary accretion disk.  In \S3,
and \S4 we solve for individual particles' orbits, and derive the
complementary set of dynamical equations for a collection of particles
in the form of low-order moments of the Boltzmann equation (Boltzmann, 1872). 
In \S3, we
focus only on small particles, which have sizes smaller than the
mean-free-path of the gas molecules, whereas in \S4, we develop
analogous equations for particles with sizes larger than the
mean-free-path of gas.  These initial calculations are purely
analytic; in order to perform them, we need to assume separation of
the radial and normal-to-the-disk variables, as well as a constant
background gas density. In \S5, we check the validity of these
assumptions with a complete numerical calculation of the particles'
orbits in a simple-to-use model for the gas nebula. We compare both
the results for individual orbits and for the moments (typically, the
average velocity and average velocity dispersion) obtained from the
Boltzmann averaging procedure.  We also consider the spatial and
temporal evolution of a given size-distribution of particles in \S6
and show that particles with a narrow selected size-range sediment
most rapidly toward the mid-plane.  Finally in \S7, we summarize our
results and discuss their implications.

\section{Gas drag laws and particle motion}

In this section, we briefly recall the effect of gas drag and
gravitational forces on individual particle trajectories. 

\subsection{Drag force description}

In dust-particle number-density regimes where particle 
collisions are negligible, the trajectory of a particle is given by
\begin{equation}
{\bf r}'' = - \grad \Phi - \frac{1}{\mp} \Fd \mbox{   ,   }
\label{eq:firsteq}
\end{equation}
where $q'$ denote derivatives of $q$ with respect to time, $\br$ is
the position vector of the particle, $\Phi$ is the externally imposed
gravitational potential, $\mp$ is the mass of the particle and $\Fd$
is the drag force between the particle and the gas. The amplitude of
the drag force depends on the size of the particle $s$ compared to the
mean free path of the gas $\lambda$ and one typically distinguishes
two regimes, $s \ll \lambda$ (Epstein regime) and $s \gg \lambda$
(Stokes regime).

In the small particle limit, {\it i.e.}  the Epstein regime, the drag
is caused by the thermal agitation of the gas and is proportional to
the velocity of the particle relative to the gas:
\begin{equation}
\Fd = m_{\rm p} \frac{\rho}{\rhos} \frac{c}{s} (\dot{\br} - \bv_{\rm
g}) \mbox{ when } s \ll \lambda \mbox{ , }
\end{equation}
where $\bv_{\rm g}$ is the gas velocity, $\rho$ is the local density
of the gas, $c$ is the local sound speed and $\rho_s$ is the solid
(i.e. internal) density of the particle. In a standard solar nebula,
this formula is typically valid up to decimeter-sized particles at 1
AU.

In the other limit, large particles see the gas as a fluid, and experience
 a drag force through the laminar or turbulent wake that they create as they move through the gas. This is the Stokes regime. Whipple (1972) reported that
the drag force on a sphere is
\begin{equation}
\Fd = m_{\rm p}\frac{\rho}{\rhos} \frac{C(Re)}{s} |\Delta \bv| (\dot{\br} - \bv_{\rm g}) \mbox{ when } s \gg \lambda \mbox{   ,   }
\end{equation}
where $|\Delta \bv|$ is the norm of $\dot{\br} - \bv_{\rm g}$ and $Re$
is the particle Reynolds number of the flow ($Re = 2 s|\Delta
\bv|/\nu$ where $\nu = \lambda c/3$ is the molecular viscosity of the
gas). Experimental results suggest that the drag coefficient $C(Re)$
varies like
\begin{eqnarray}
C &=& 9 Re^{-1} \mbox{   for    } Re \le 1 \mbox{   ,   } \nonumber \\
C &=& 9 Re^{-0.6 } \mbox{   for    } 1 \le Re \le 800 \mbox{   ,   } \nonumber \\
C &=& 0.165 \mbox{  for    }  800 \le Re \mbox{   .   }
\end{eqnarray}
This prescription ensures a smooth transition between the Epstein
regime and the Stokes regime for intermediate particle sizes ($s
\simeq \lambda$). Note that in the Stokes regime (for $Re \rightarrow
\infty$), the amplitude of the drag force depends on the total
velocity of the particle with respect to the gas, and not simply its
component parallel to the particle's velocity. This expresses the fact
that a strong wake caused by the particles motion in one direction
also affects even the slightest motion in another perpendicular
direction.

\subsection{Separation of vertical and radial motions}

As particles undergo gas drag, they lose angular momentum which causes
them to drift slowly inwards. For particles strongly coupled to the
gas (i.e. very small particles), this drift is slowed down by the
support provided by the gas pressure itself. For weakly coupled
particles (i.e. very large particles), the angular-momentum loss is
very small and the drift is equally slow (see the next section). There
exists an intermediate regime, however, where orbital decay can occur
very rapidly. Weidenschilling (1977) carried out a first quantitative
study of the effect of the gas drag on particles of various sizes, and
determined the inward drift velocity as a function of particle
size. He found that the maximal drift occurs for particles for which
the typical stopping timescale $t_s = |(\br' -\bv_{\rm g})|/|\Fd|$ is
of order of the orbital timescale $\ok^{-1}$, and decays very quickly
for particles much larger, or much smaller than that.  As a result,
unless the particle is of intermediate size, there will be a marked
separation between the dynamical timescale and the orbital decay
timescale.

Moreover, there is also a marked separation between radial and
vertical length scales; indeed, numerical integration of the orbits of
particles in a gaseous disk shows that the typical radial drift
velocity is of the same order (in the case of small particles) or much
smaller (in the case of large particles) than the vertical settling
velocity (see Section \ref{sec:numint_traj}). Hence in a thin disk the
particles settle to the mid-plane with very little radial excursion,
then undergo radial decay within a very thin dust layer. For this
reason it is justified to assume that the vertical motion of particles
is more or less independent of their radial motion (although not of
their radial {\it position}), and that their radial motion occurs
mostly at $z=0$. Since Weidenschilling (1977) has already derived an
analytical description of the particles motion in the radial
direction, we shall concentrate on the problem of vertical settling. A
short derivation of the case of radial and azimuthal motion of the
particles in a gas disk is presented in the Appendix for completeness.

\section{The collective evolution of small particles}
\label{sec:eps}

\subsection{Vertical motion of small particles in the Epstein regime}

From the previous sections, we infer the vertical equation of motion
of a small dust-particle in the Epstein regime:
\begin{equation}
z'' = - \frac{\ptl \Phi}{\ptl z} - \frac{\rho}{\rhos}\frac{c}{s} z' \mbox{   ,   }
\end{equation}
where $z$ is the height above the disk. In a central potential, the
gravitational force close to the mid-plane can be rewritten as
$-\ok^2(r) z$ where $\ok(r)$ is the Keplerian angular velocity at the
radius considered. We re-normalize the time variable to the orbital
timescale $t_o=\ok^{-1}$ at the radial position considered, and the
distances to Astronomical Units (AU); we deduce the non-dimensional
equation
\begin{equation}
\ddot{z} = - z - \mu \dot{z} \mbox{   ,   }
\label{eq:epseq}
\end{equation}
where $\mu = \ok^{-1}(r) (\rho(r)/\rhos)(c(r)/s)$. In terms of this
normalization, the magnitude of $\mu$ in this regime is typically much
larger than unity, which means that the stopping time $1/\mu$ is much
smaller than the orbital time. The solution to equation (\ref{eq:epseq}) is
straightforward, and the trajectory of a particle initially positioned
at height $z_i$ with initial velocity $ w_i$ is
\begin{eqnarray}
\zp(\mu,t;z_i,w_i) &=& \frac{1}{2}\left[ \left(z_i + \frac{2w_i + \mu
z_i}{\sqrt{\mu^2-4}}\right) \exp\left(\frac{-\mu +
\sqrt{\mu^2-4}}{2}t\right) \right. \nonumber \\ &+& \left. \left(z_i -
\frac{2w_i + \mu z_i}{\sqrt{\mu^2-4}}\right) \exp\left(\frac{-\mu -
\sqrt{\mu^2-4}}{2}t\right) \right] \mbox{ .   } 
\label{eq:epstraj}
\end{eqnarray}
Note that for $\mu \gg 1$ this expression
simplifies to
\begin{equation}
\zp(\mu,t;z_i,w_i) = \left( z_i + \frac{w_i}{\mu} \right) e^{-t/\mu} -
\frac{w_i}{\mu} e^{-\mu t} \mbox{   .   }
\end{equation} 
The second term of this expression describes the rapid deceleration of
the particle by the gas drag, the first represents a slow settling
toward the mid-plane. Let us recast this equation into
\begin{equation}
\zp(\mu,t;z_i,w_i) = \alpha(\mu,t) z_i + \beta(\mu,t) w_i \mbox{   ,   }
\label{eq:5}
\end{equation}
which defines the functions $\alpha$ and $\beta$ uniquely. The
instantaneous vertical velocity of these particles is then given by
\begin{equation}
\wp(\mu,t;z_i,w_i) = \dot{z}_{\rm p} = \dot{\alpha} z_i + \dot{\beta} w_i
\mbox{ .  }
\end{equation}
Note that for $\mu \simeq 1$ (which describes the intermediate
parameter range between the Epstein regime and the Stokes regime) the
particle motion contains an oscillatory component decaying
exponentially on timescale $\mu/2$.

\subsection{Boltzmann description and continuum equations}

Given a set of particles at initial positions $z_i$ and with initial
velocities $w_i$, the Boltzmann distribution function of these
particles is
\begin{equation}
f(z,w,t) = \sum_i m_i \delta(z-\zp(\mu,t;z_i,w_i))
\delta(w-\wp(\mu,t;z_i,w_i)) \mbox{ . }
\end{equation}
One can also take the continuum limit of this description in the case of a
very large number of particles: 
\begin{eqnarray}
f(z,w,t) = \int \dd z_i \int \dd w_i && \rho_i(z_i) g(w_i,z_i)
\delta(z-\zp(\mu,t;z_i,w_i)) \nonumber \\ && \cdot
\delta(w-\wp(\mu,t;z_i,w_i)) \mbox{   ,   }
\end{eqnarray}
where $\rho_i$ is the initial particle density distribution, and $g$
is the initial velocity distribution of particles. The freedom in the
relative choices of the initial distribution functions $\rho_i$ and
$g$ is lifted by requiring that for all $z_i$
\begin{equation}
\int \dd w_i \mbox{  }g(w_i,z_i) = 1 \mbox{   .   }
\end{equation}

The mass density of the particles is obtained by integrating $f$ over
all possible velocities:
\begin{align}
\rhop(z,t) =& \int \dd w f(z,w,t) \nonumber  \\ =& \int \dd z_i \int \dd w_i
\rho_i(z_i) g(w_i,z_i) \delta(z-\zp(\mu,t;z_i,w_i)) \mbox{ .  }
\label{eq:1}
\end{align}
Substitution of the equations for the particle trajectories into
(\ref{eq:1}) and integration over all possible initial positions
yields
\begin{equation}
\rhop(z,t) = \frac{1}{\alpha} \int \dd w_i \rho_i\left(\frac{z-\beta
w_i}{\alpha}\right) g\left(w_i,\frac{z-\beta w_i}{\alpha}\right)\mbox{   .   } 
\label{eq:2}
\end{equation}

The first moment of the Boltzmann function is the average vertical
velocity:
\begin{align}
\rhop(z,t) \overline{w} =& \int \dd w w f(z,w,t) \\ =&
\int \dd z_i \int \dd w_i \rho_i(z_i) g(w_i,z_i) \wp
\delta(z-\zp(\mu,t;z_i,w_i)) \mbox{   .   } \nonumber 
\end{align}
The same manipulations yield
\begin{equation}
\rhop \overline{w} = \int \frac{\dd w_i}{\alpha}
\rho_i\left(\frac{z-\beta w_i}{\alpha}\right) g\left(w_i,\frac{z-\beta
w_i}{\alpha}\right) \left( \dot{\alpha} \frac{z-\beta w_i}{\alpha} +
\dot{\beta} w_i \right)
\label{eq:3}
\end{equation}

Let us assume for a short while that the distribution of the initial
velocities is independent of height above the mid-plane. Then we can
verify easily that the quantities determined above in equations
(\ref{eq:2}) and (\ref{eq:3}) satisfy the standard continuity equation
\begin{equation}
\frac{\ptl \rhop}{\ptl t} + \frac{\ptl }{\ptl z} (\rhop \overline{w}) = 0 \mbox{   ,   }
\end{equation}
regardless of the form of 
the ``trajectory functions'' $\alpha$ and $\beta$.

However, the original collisionless Boltzmann equation (Boltzmann, 1872) 
describes the evolution of
the distribution function $f$ in phase space through
\begin{equation}
\frac{\ptl f}{\ptl t} + w \frac{\ptl f}{\ptl z} + \dot{w} \frac{\ptl
f}{\ptl w} = \Gamma \mbox{   ,   }
\end{equation}
where $\Gamma$ is a collision term that reproduces the interaction of
the particles with themselves and with the surrounding medium. If we
integrate the Boltzmann equation with respect to the velocity space, we get
\begin{equation}
\frac{\ptl \rhop }{\ptl t} +\frac{\ptl}{\ptl z}(\rhop \overline{w}) +
\int \dd w \dot{w} \frac{\ptl f}{\ptl w} = \int \Gamma \dd w \mbox{   .   }
\end{equation}
Substitution of the equation of motion equation (\ref{eq:epseq}) 
into the last term of the left-hand-side
provides an expression for the interaction term $\Gamma$ as a
condition for the mass continuity equation to be satisfied:
\begin{equation}
\Gamma = \mu f \mbox{   .   }
\end{equation}
This result can be generalized in the case where $g$ depends on $z_i$.

Subtracting the mass continuity equation from the first moment of the
Boltzmann equation reveals the importance of the particle velocity
correlations $\overline{w^2}$
\begin{equation}
\rhop \frac{\ptl \overline{w}}{\ptl t} +\rhop \overline{w} \frac{\ptl
\overline{w}}{\ptl z} = - \rhop \frac{\ptl \Phi}{\ptl z} - \mu \rhop
\overline{w} - \frac{\ptl }{\ptl z} (\rhop \sigma^2) \mbox{   ,   }
\end{equation}
where $\sigma^2 = \overline{w^2} - \overline{w}^2$ represents the
particles' velocity dispersion function. This equation is very similar
to that of a standard fluid, with the following caveats: (i) it
contains an explicit drag term in the form of $-\mu \overline{w}$, and
(ii) the very last term, which usually represents the pressure term in
a standard fluid\footnote{this term is related to thermal elastic
collisions of the gas particles in a standard fluid, and tends to
depend only on the local density and temperature of the fluid}, must
be written explicitly as a function of known quantities. We expect
this term to be null in the case of collisionless particles 
with negligible relative velocity with respect to the gas; in order
to double check this conjecture, we now evaluate $\sigma^2$
explicitly.

Let us first evaluate $\overline{w^2}$:
\begin{equation}
\rhop \overline{w^2} = \int \frac{\dd w_i}{\alpha}
\rho_i \hspace{-0.5pt} \left(\frac{z-\beta w_i}{\alpha}\right) g\left(w_i,\frac{z-\beta
w_i}{\alpha}\right) \hspace{-2pt} \left( \dot{\alpha} \frac{z-\beta w_i}{\alpha} +
\dot{\beta} w_i \right)^2 
\end{equation}

The behavior of $\sigma^2$ at short times (i.e. for times shorter than
the stopping time $t_{\rm s} = 1/\mu$) is difficult to extract
analytically while keeping the functions $\rho_i$ and $g$
unprescribed. However, for large $\mu$ and large times one can
simplify the expressions for the functions $\alpha$ and $\beta$ to
\begin{eqnarray}
\alpha &=& \exp(-t/\mu) + O(\exp(-\mu t)) \mbox{   ,   } \nonumber \\
\beta &=& \frac{1}{\mu} \left(\exp(-t/\mu) - \exp(-\mu t)\right) \mbox{   .   }  
\end{eqnarray}
In that case,
the expressions for the first and second velocity moments simplify
largely to
\begin{eqnarray}
\overline{w} &=& - \frac{z}{\mu} + O(\exp(-\mu t)) \mbox{   ,   } \nonumber \\ 
\overline{w^2} &=& \frac{z^2}{\mu^2} + O(\exp(-\mu t)) \mbox{   ,   }
\end{eqnarray}
and one can show that provided $g$ is an even function of velocity
\begin{equation}
\sigma^2 = O(\exp(-2\mu t)) \mbox{   .   }  
\end{equation}
This result is expected: any initial velocity dispersion is quickly
damped out by the surrounding gas on the short stopping timescale
$t_{\rm s}$. Note that in the case where $\rho_i$ and $g$ are uniformly distributed, this expression is also valid for very short times.
The long-term evolution of the collection of particles
becomes a slow collective settling, on a timescale $\mu$. Therefore,
in the Epstein case, one can approximate the particles as a fluid with
a linear drag force and zero pressure.

\section{The collective evolution of large particles}
\label{sec:sto}

The main drag effect of the gas on a large particle occurs through the
formation of a wake behind the particle. If the velocity of the
particle relative to the gas is small enough, the wake is laminar and
the drag force is a linear function of the relative velocity. In that
case, the particle trajectory is actually the same as that obtained in
the case of the Epstein regime, and the results of the previous
section apply. However, as the relative velocity between the particle
and the gas increases, the wake becomes turbulent and the amplitude of
the drag force becomes a power law of the particle's Reynolds number.

As a result, though the component of the drag force in any direction
is still proportional to the relative velocities of the particle and
the gas in that same direction, its amplitude also depends on the
total relative velocity of the particle with respect to the
fluid. Hence the vertical component of the drag force is:
\begin{equation}
\Fd \cdot \ez = - \mp \frac{\rho(r)}{\rhos} \frac{C(Re)}{s} \sqrt{ r'^2 + z'^2 + (r\theta' - \vg)^2 } z' \mbox{   ,   }
\label{eq:realF}
\end{equation}
if one assume the only important component of the gas velocity is in
the azimuthal direction. This formula can in principle be used only in
conjunction with a complete evaluation of the particle's
orbit. However, one can assume that the largest typical contrasts in
velocity between particle and gas occurs in the vertical direction,
where the particles oscillate across the mid-plane with an epicyclic
frequency whereas the gas is more-or-less stationary (so that $z'^2
\gg (r\theta' - \vg)^2$ and $z'^2 \gg r'^2$). In that case, the
expression for the vertical component of the drag force reduces to
\begin{equation}
\Fd \cdot \ez = - \mp \frac{\rho(r)}{\rhos} \frac{C(Re)}{s} |z'| z' \mbox{   .   }
\label{eq:badF}
\end{equation}

Note that since the typical velocity of the particles across the
mid-plane is of order of $\Delta \Omega_{\rm K}$ where $\Delta$ is the
typical height of particles above the disk, the asymptotic Stokes
regime occurs for particles of size
\begin{equation}
\frac{s}{\lambda} = \frac{Re_{\rm c}}{6} \frac{r}{\Delta}
\frac{\vk}{c} \mbox{ , }
\end{equation}
where $Re_{\rm c} = 800$. This condition corresponds for example to 10
m size particles at $r=1$ AU and $\Delta = 1/10$th of the disk scale
height in a standard solar nebula.

\subsection{Particles' trajectories}

Using the same normalizations as in the Epstein regime, the particle's
vertical equation of motion can then be written as
\begin{equation}
\ddot{z} = -z - \mu |\dot{z}| \dot{z}\mbox{   ,   }
\label{eq:stoeq}
\end{equation}
where $\mu = C(Re) (R/s) (\rho/\rhos)$ and $R=1$ AU. In the asymptotic
Stokes regime, $C(Re)$ is simply a constant, and $\mu $ is typically
of order of unity, or smaller.

This equation can be solved exactly through the introduction of the
quantity $w= \dot{z}$. On the positive branch (when $w>0$) and
negative branch ($w<0$) the solutions are, respectively,
\begin{eqnarray}
w^2_+ &=& A \exp(-2\mu z) - \frac{z}{\mu} + \frac{1}{2\mu^2} \mbox{ ,
}\nonumber \\ w^2_- &=& B \exp(2\mu z) + \frac{z}{\mu} +
\frac{1}{2\mu^2}\mbox{ .  }
\end{eqnarray}
As the particle oscillates about the mid-plane, it follows one branch
or the other. The constants $A$ and $B$ are determined by matching the
first branch to the boundary conditions and the successive alternative
ones to each other at the turning point (i.e. when $w=0$). For
instance, the trajectory of a particle starting from rest at $z=z_0$
above the mid-plane satisfies the equation
\begin{equation}
w_-^2 = \dot{z}^2 = \left(-\frac{z_0}{\mu} - \frac{1}{2\mu^2}\right)
\exp(2\mu z - 2\mu z_0) + \frac{z}{\mu} + \frac{1}{2\mu^2} \mbox{   .   }
\end{equation}
It accelerates toward the mid-plane, then decelerates on the other
side until it reaches a turning point $z_1$ which satisfies
\begin{equation}
0 = \left(-\frac{z_0}{\mu} - \frac{1}{2\mu^2})\right) \exp(2\mu z_1 -
2\mu z_0) + \frac{z_1}{\mu} + \frac{1}{2\mu^2}\mbox{   .   }
\end{equation}
These solutions indicate the existence of two regimes. When the
initial height $z_0$ is much larger than the scale height $1/\mu$ the
turning point $z_1$ is roughly equal to $1/2\mu$ regardless of the
height from which it originated. This pattern corresponds to a rapid
stopping phase, when the particle quickly drops toward the mid-plane.
When the amplitude of $z_0$ is reduced much below $1/\mu$, the turning
point $z_1$ is roughly equal to $-z_0$, which corresponds to the
oscillatory phase, when the particle follows an epicyclic motion about
the mid-plane with a slowly decaying amplitude.

\subsection{The very rapid stopping phase}

The stopping phase corresponds to the limit where $z$ is much larger
than the scale height $1/\mu$. Note that this situation may not always
occur since in the Stokes regime, $\mu \le 1$. This condition
corresponds to particles which start at a considerable distance above
or below the disk itself. Nonetheless, in this case the solutions
simplify to:
\begin{equation}
w_-^2 \simeq \frac{z}{\mu} \equiv \dot{z} \simeq -
\frac{z^{1/2}}{\mu^{1/2}} \mbox{   ,   }
\end{equation} 
(for the downward branch) and this differential equation 
can easily be solved with 
\begin{equation}
\zp(t) = \left(z_0^{1/2} - \frac{t}{2\mu^{1/2}} \right)^2 \mbox{   .   }
\end{equation}
Note that the particle velocity in that case varies linearly with
time:
\begin{equation}
\wp(t) = - 2 \left(\frac{z_0}{\mu}\right)^{1/2} + \frac{t}{2\mu} \mbox{   ,   }
\end{equation}
so that the typical duration of the settling phase is $t_s = 2
\mu^{1/2} z_0^{1/2}$. Finally, we also note that for a particle
starting from $z_0$ above the disk, for example, the lower turning
point is more or less independent of $z_0$ and is roughly equal to
$z_1 = -1/2\mu$.

\subsection{The slowly decaying oscillating phase}

The oscillating phase corresponds to the limit $z \ll 1/\mu$. In that
case, the equation for the trajectory of the particle (in the downward
branch, for example) simplifies to
\begin{equation}
w^2_- = \dot{z}^2 = 2 z_0 (z_0 - z) - (1+2\mu z_0)(z_0-z)^2 \mbox{   .   }
\end{equation}
This first order differential equation can be solved with conventional
substitutions to yield
\begin{equation}
\zp(t) = z_0 - \frac{z_0}{1+2\mu z_0} \left[ 1 - \cos \left( (1+ 2\mu
z_0)^{1/2} t\right) \right] \mbox{   .   }
\end{equation}
The lower turning point is then simply
\begin{equation}
z_1  = z_0 - \frac{2 z_0}{1+2\mu z_0} \mbox{   .   }
\end{equation}
Starting from that point, the upward branch is then obtained through a
similar analysis and becomes
\begin{equation}
\zp(t) = z_1 - \frac{z_1}{1-2\mu z_1} \left[ 1 - \cos \left( (1- 2\mu
z_1)^{1/2} t\right) \right] \mbox{   .   }
\end{equation}
The following upper turning point is $z_2$ such that
\begin{equation}
z_2 = z_1 - \frac{2 z_1}{1-2\mu z_1} \mbox{   .   } 
\end{equation}
These equations show that the amplitude $\gamma(t)$ of the oscillation slowly
decays with time, and can be approximated by
\begin{equation}
\gamma(t)\simeq \frac{3\pi}{ 4\mu t + K} \mbox{   ,   }
\label{eq:4}
\end{equation}
where $K$ is a constant which is determined in such a way as to
satisfy the initial conditions. If the particle is released from rest
from $z=z_0$ at $t=0$ (with $z_0 \ll 1/\mu$), $K = 3\pi/z_0$.
Moreover the frequency of the oscillation also varies with time, and
converges to the epicyclic frequency.

\begin{figure}[t]
\plotone{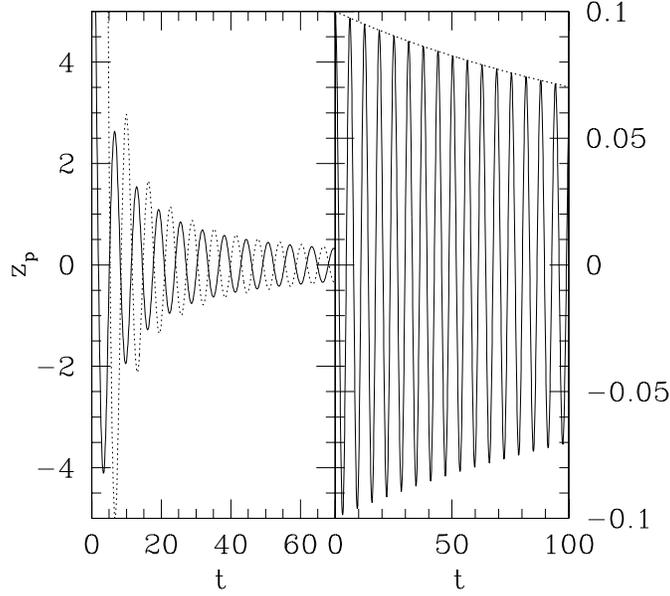}
\caption{{\it Left panel:} Trajectories of two particles as integrated
numerically from the differential equation (\ref{eq:stoeq}). The
particles were released from rest from heights 10$r$ (solid line) and
100$r$ (dotted line), in the case where $\mu = 0.1$. Note that the
first lower turning point is always located around $-1/2\mu$ and that
the oscillation amplitude is independent of the point of release in
the following oscillating period. However, depending on the initial
condition, a phase difference can exist between two trajectories. {\it
Right panel: } Trajectories of a particle released from rest from
height 0.1 above the disk, with $\mu = 0.1$. The particle oscillates
around the mid-plane with an amplitude that decreases algebraically
with time; the dotted line was drawn according to the amplitude
given by equation (\ref{eq:4}).}
\label{fig:veloc}
\end{figure}

To recapitulate briefly, we find that after a brief stopping phase on
a timescale proportional to $\mu^{1/2}$, the particle oscillates
around the mid-plane with frequency which is close to the epicyclic
frequency, and with an amplitude that decays slowly in time roughly as
given by equation (\ref{eq:4}). These solutions are illustrated in
Figure \ref{fig:veloc}.

\subsection{Continuum equations in the Stokes regime}

The true particle trajectory in the asymptotic Stokes regime cannot be
expressed analytically in any simple way. As in the Epstein case, we
will only look at long-term behavior. In that case, we can approximate
the trajectory of a particle by
\begin{equation}
\zp(t,\mu;K_i,\varphi_i) = \gamma(\mu,t;K_i) \cos (t+ \varphi_i) \mbox{   ,   }
\label{eq:stotraj}
\end{equation}
where $K_i$ and $\varphi_i$ are complicated functions of the initial
position and velocity of the particle. For simplicity, we will
consider the asymptotic limit for times $4\mu t >> K_i$. In this
limit, equation (\ref{eq:stotraj}) with $\gamma = 3\pi/4\mu t$ is a
good approximation, since the amplitude of the particle trajectory
becomes independent of the point of release (see
Fig. \ref{fig:veloc}). Let us also assume for the purpose of this work
that the phase simply has a given distribution function $g(\varphi)$
so that Boltzmann function for the Stokes regime is
\begin{eqnarray}
f(z,w,t) &=& \int \dd z_0 \rho_i (z_0) \int_0^{2\pi} \dd \varphi
g(\varphi) \delta(z-\zp(\mu,t;\varphi)) \nonumber \\ &=& \Sigma_{\rm
p} \int \dd \varphi g(\varphi) \delta(z-\zp(\mu,t;\varphi)) \mbox{   ,   }
\label{eq:fsto}
\end{eqnarray}
where $ \Sigma_{\rm p}$ is the column density of particles, and the
normalization of $g$ is chosen such that $\int_0^{2\pi} \dd \varphi
g(\varphi) = 1$. The function $g$ is periodic with period $2\pi$ but
the initial distribution of particles can be described with $0 \le
\varphi \le 2\pi$. Using the well-known relation
\begin{equation}
\delta(h(x)) = \sum_i \delta(x-x_i) \left|\frac{\dd h}{\dd x}(x=x_i)
\right|^{-1} \mbox{ where } h(x_i) = 0 \mbox{   ,   }
\end{equation}
we can determine the particle mass density profile,
\begin{eqnarray}
\rhop(z,t) = && \frac{\Sigma_{\rm p}}{\gamma} \calH (\gamma - |z|)
\left(1-\frac{z^2}{\gamma^2}\right)^{-1/2} \nonumber \\ && \cdot
\left[ g\left( \cos^{-1}(z/\gamma) - t \right) + g\left(-
\cos^{-1}(z/\gamma) - t \right) \right] \mbox{   ,   }
\end{eqnarray}
where $\calH$ is a Heaviside function and the function $\cos^{-1}$
takes values in the interval $[0,\pi]$.
 
As in the Epstein case, we can compare the fluid equation for mass
conservation to the first moment of the Boltzmann equation and obtain
an expression for the collision term in the Stokes regime:
\begin{equation}
\Gamma = 2\mu |w| f \mbox{  .   }
\end{equation}

Similar operations with the second moment of the Boltzmann equation
yields the equivalent of the fluid equation of motion,
\begin{equation}
\rhop \frac{\ptl \overline{w}}{\ptl t} +\rhop \overline{w} \frac{\ptl
\overline{w}}{\ptl z} = - \rhop \frac{\ptl \Phi}{\ptl z} - \int \mu f
|w|w \mbox{  }\dd w - \frac{\ptl }{\ptl z} (\rhop \sigma^2) \mbox{  .   }
\end{equation}
It is extremely tempting to write the drag term as 
\begin{equation}
\int \mu f |w|w \mbox{  }\dd w = \mu \rhop |\overline{w}| \overline{w} \mbox{  .   }
\label{eq:wrongdrag}
\end{equation}
However, a proper evaluation of the left-hand-side integral shows that
this cannot be done formally, and the complete expression for the drag
force term should be kept. From equations (\ref{eq:stotraj}) and
(\ref{eq:fsto}),
\begin{eqnarray}
&& \int f |w|w \mbox{ }\dd w = \frac{\Sigma_{\rm p}}{\gamma}
\left(1-\frac{z^2}{\gamma^2}\right)^{-1/2} \calH (\gamma - |z|) \\ &&
\cdot \left[g(\cos^{-1}(z/\delta) - t ) \left(\frac{z
\dot{\gamma}}{\gamma} - \gamma \sqrt{1-\frac{z^2}{\gamma^2}} \right)
\left|\frac{z \dot{\gamma}}{\gamma} - \gamma
\sqrt{1-\frac{z^2}{\gamma^2}} \right| \right. \nonumber \\ &+& \left.
g(- \cos^{-1}(z/\delta) - t ) \left( \frac{z \dot{\gamma}}{\gamma} +
\gamma \sqrt{1-\frac{z^2}{\gamma^2}} \right) \left|\frac{z
\dot{\gamma}}{\gamma} + \gamma \sqrt{1-\frac{z^2}{\gamma^2}} \right|
\right] \mbox{ .  } \nonumber
\end{eqnarray}
In the case where $g$ is uniformly distributed, this expression
reduces to
\begin{eqnarray}
\int f |w|w \dd w &=& \frac{\rhop}{2} \left[\left(\frac{z \dot{\gamma}}{\gamma} - \gamma \sqrt{1-\frac{z^2}{\gamma^2}} \right) \left| \frac{z \dot{\gamma}}{\gamma} - \gamma \sqrt{1-\frac{z^2}{\gamma^2}} \right| \right. \nonumber \\ &+& \left. \left(\frac{z \dot{\gamma}}{\gamma} + \gamma \sqrt{1-\frac{z^2}{\gamma^2}} \right) \left| \frac{z \dot{\gamma}}{\gamma} + \gamma \sqrt{1-\frac{z^2}{\gamma^2}} \right|\right] \mbox{  ,   }
\end{eqnarray}
whereas the simpler expression $\rhop |\overline{w}| \overline{w}$ in
equation (\ref{eq:wrongdrag}), which is rewritten $\rhop (z
\dot{\gamma}/{\gamma}) |z \dot{\gamma}/{\gamma}|$ in that case, would
be wrongly applied.

As before, we are interested in evaluating the velocity dispersion
$\sigma^2$. We need to calculate
\begin{equation}
\frac{\rhop \overline{w}}{\Sigma_{\rm p}} = \hspace{-2.5pt} \int_0^{2\pi}\hspace{-2.5pt}  \dd \varphi
g(\varphi) \delta(z-\gamma \cos(\varphi+t)) \left( \dot{\gamma}
\cos(t+\varphi) - \gamma \sin(t+\varphi)\right) \mbox{ .  }
\end{equation}
In the asymptotic limit 
\begin{align}
\frac{\rhop \overline{w}}{ \Sigma_{\rm p}} =&  \frac{\calH (\gamma -
|z|) }{\gamma} \left[\left(
  \frac{\dot{\gamma}}{\gamma}\left(1-\frac{z^2}{\gamma^2}\right)^{-1/2}
  z - \gamma \right) g\left(\cos^{-1}(z/\delta) - t\right) \right. \nonumber \\
  +& \left. \left(
  \frac{\dot{\gamma}}{\gamma}\left(1-\frac{z^2}{\gamma^2}\right)^{-1/2}
  z + \gamma\right) g\left(-\cos^{-1}(z/\delta) - t \right) \right] \\ =& \frac{\rhop}{\Sigma_{\rm p}} \frac{\dot{\gamma}}{\gamma}z - \calH (\gamma - |z|)
\nonumber \\ \cdot & \left[ g\left(\cos^{-1}(z/\delta) - t \right) - g\left(-\cos^{-1}(z/\delta) - t \right)
  \right] \mbox{ .  } \nonumber
\end{align}

Similarly, one can show that 
\begin{eqnarray}
\rhop \overline{ww} = \Sigma_{\rm p} \int_0^{2\pi} \dd \varphi &&
g(\varphi) \delta(z-\gamma \cos(\varphi+t)) \nonumber \\ && \cdot
\left( \dot{\gamma} \cos(t+\varphi) - \gamma \sin(t+\varphi)\right)^2
\mbox{ , }
\end{eqnarray} 
so that in the asymptotic limit
\begin{eqnarray}
\rhop \overline{ww}& =& \rhop \left(\frac{z^2
\dot{\gamma}^2}{\gamma^2} + (\gamma^2 - z^2)\right) - 2 \Sigma_{\rm
p}\calH (\gamma - |z|)\dot{\gamma}z \nonumber \\ && \cdot \left[
g(\cos^{-1}(z/\delta) - t ) - g(-\cos^{-1}(z/\delta) - t ) \right]
\mbox{ .  }
\end{eqnarray}

In the case where the initial distribution function $g$ is uniform, we
can simplify these expressions greatly and show that for $|z| \le
\gamma$ (which corresponds to the thickness of the dust layer) then
\begin{equation}
\sigma^2 = (\gamma^2 - z^2) \calH (\gamma -|z|) \mbox{  .   }
\end{equation}
This expression means that within the dust disk, the particle velocity
dispersion remains important at all times. The velocity dispersion
mimics the effect of a pressure term which effectively slows down the
settling. In this simplified case where $g(\varphi)$ was taken to be a
uniform distribution, it is actually possible to relate $\sigma$ to
intrinsic large-scale properties of the system. However in the more
general case where $g(\varphi)$ is not a uniform distribution there
exists no simple relationship between $\sigma^2$ and the local density
of the gas as there would normally be in a standard fluid. Instead,
this mock-pressure terms would depend in a complicated manner on the
initial configurations of the particles in phase-space. This prevents
any further progress, at this stage, in the use of a fluid description
for these large particles.

\section{Numerical calculations}
\label{sec:numint}

With this simplified analytical approach, we have explored what
effects the gas drag may have on particles of various sizes. Three
principal approximations were performed:
\begin{itemize}
\item we neglected the radial motion of the particles, and therefore
implicitly assumed that it is possible to perform a separation of the
variables and of the equations of motion into individual components;
\item we neglected the vertical variation of $\mu$ and used a simplified expression for the gravity,
\item we neglected the contribution of the azimuthal motion of the
particles through the gas in the calculation of the gas drag in the
Stokes regime. This approximation effectively underestimates the gas
drag.
\end{itemize}

In order to check the validity of these approximations, we begin by
integrating the orbits of particles in a realistic gaseous
protoplanetary nebula, with the complete expression for the gas drag
force. We compare them to the corresponding analytical expressions. We
then release various single-size sets of 10,000 particles in the disk
from a small region and follow their evolution in time. At a given
time $t$ we then perform the Boltzmann averaging procedure and compare
the results with those obtained in Sections \ref{sec:eps} and
\ref{sec:sto}.

\subsection{Numerical Integration: method and model parameters}
\label{sec:numint_desc}

The equation of motion of a particle in a gaseous accretion disk is
given by equation (\ref{eq:firsteq}). Expanding this equation into its
components in a cylindrical coordinate system $(r,\theta,z)$ yields
\begin{eqnarray}
r'' &=& r \theta'^2 - \frac{GMr}{(r^2 + z^2)^{3/2}} - \frac{\Fd}{m_{\rm p}} \cdot \er \mbox{  ,   } \nonumber \\
\theta''&=& -2\frac{r'}{r} \theta' - \frac{\Fd}{m_{\rm p}}\cdot \etheta \mbox{  ,   }\nonumber  \\ 
z'' &=& - \frac{GMz}{(r^2 + z^2)^{3/2}}  - \frac{\Fd}{m_{\rm p}} \cdot \ez\mbox{  ,   }
\end{eqnarray}
where for small particles, we use the Epstein drag force expression
\begin{equation}
\frac{\Fd}{m_{\rm p}} = \frac{\rho}{\rhos} \frac{c}{s} (r'\er + (r \theta' - \vg) \etheta + z' \ez) \mbox{  ,   }
\end{equation}
and for very large particles, we use the Stokes drag force
\begin{equation}
\frac{\Fd}{m_{\rm p}} =   \frac{\rho}{\rhos} \frac{C(Re)}{s} \sqrt{r'^2 + (r \theta' - \vg)^2 + z'^2}  (r'\er + (r \theta' - \vg) \etheta + z' \ez) \mbox{  .   }
\end{equation}

We normalize these expressions using $R = 1AU = 1.5 \times 10^{13}$cm
as unit distance, and $\Omega^{-1}_{\rm K}(R)= 2.0 \times 10^{-7}$s as
a unit time. This yields
\begin{eqnarray}
\ddot{r} &=& r \dot{\theta}^2 - \frac{r}{(r^2 + z^2)^{3/2}} - \mu(r,z) \dot{r}
 \mbox{  ,  } \nonumber  \\
\ddot{\theta}&=& -2\frac{\dot{r}}{r} \dot{\theta} - \mu(r,z) \left(\dot{\theta} - \frac{\vg(r,z)}{r v_{\rm K}(R)}\right) \mbox{  ,  } \nonumber  \\ 
\ddot{z} &=& - \frac{z}{(r^2 + z^2)^{3/2}}  - \mu(r,z) \dot{z} \mbox{  ,  }
\end{eqnarray}
where $v_{\rm K}(R)$ is the linear azimuthal Keplerian velocity at $R=1AU$ and where in the Epstein regime,
\begin{equation}
\mu_{\rm E}(r,z) = \frac{\rho(r,z)}{\rhos} \frac{c(r,z)}{s \Omega_{\rm K}(R)} \mbox{  ,  }
\end{equation}
and in the Stokes regime
\begin{equation}
\mu_{\rm S}(r,z) = \frac{\rho(r,z)}{\rhos} C(Re)\frac{R}{s} \mbox{  .  }
\end{equation}
The numerical integration the particles' trajectories in this model is
performed using a fourth order Range-Kutta integrator.

The quantities related to the disk structure (as the gas density
$\rho$, the gas velocity $v_{\rm g}$ and the sound speed $c$) are
derived from the protoplanetary nebula model of Supulver \& Lin
(2000). In this model, the gas follows a polytropic equation of state
with
\begin{equation}
p = K \rho^2 \mbox{    and   } c = \sqrt{\gamma K \rho}   \mbox{  ,  }
\end{equation}
where $p$ is the gas pressure, with $K = 6.9 \times 10^{20}$ dyn cm$^4$
g$^{-2}$, $\gamma = 1.4$ for an ideal diatomic gas (composed of pure
molecular hydrogen for example) and
\begin{equation}
\rho(r,z) = 8.5 \times 10^{-12} \left(\frac{r}{6}\right)^{-3/4}
\left(1-z^2/H^2(r)\right) \mbox{ g cm}^{-3} \mbox{ , }
\end{equation}
where $r$ is in AU and $H^2(r) = 34\times 10^{-12} K
(r/6)^{-3/4}\Omega_{\rm K}(r)^{-2} R^{-2} $ is the square of the disk
scale height (in the dimensionless units). Finally, the dimensionless
gas velocity is
\begin{eqnarray}
\frac{v_{\rm g}(r)}{v_{\rm K}(R)} &=& \frac{ v_{\rm K}(r)}{v_{\rm
K}(R)} + \frac{1}{2 \rho}\frac{\ptl \rho}{\ptl r} \frac{1}{v_{\rm
K}(R) \Omega_{\rm K}(r)} \\ &=& r^{-1/2} - \frac{8.5 \times 10^{-12} K
R}{GM} \frac{3}{4} 6^{3/4} r^{-1/4} = r^{-1/2} - \epsilon r^{-1/4}
\mbox{ , } \nonumber
\label{eq:subk}
\end{eqnarray}
if we ignore the variation of the gas density scale height with
radius.  This expression defines $\epsilon$ uniquely; with the
numerical values of $K$, $G$, given above and using $M = 2\times
10^{33} g$ we have $\epsilon = 1.8 \times 10^{-3}$. This expression
describes how gas pressure reduces the effective gravity on the gas;
this results in the slightly sub-Keplerian character of the gas
velocity as found in equation (\ref{eq:subk}).

\subsection{Numerical integration: particles trajectories}
\label{sec:numint_traj}

\subsubsection{Small particles}

\begin{figure}
\plotone{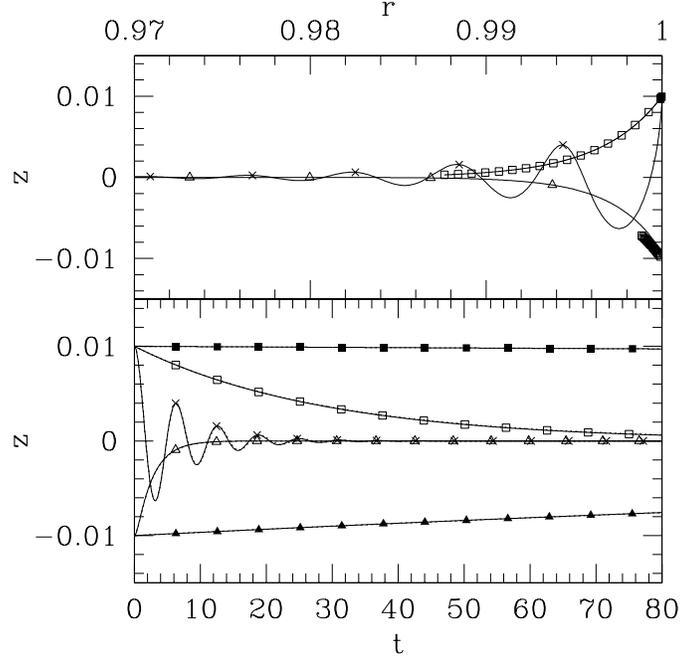}
\caption{{\it Top figure}: trajectories of small particles in the
$(r,z)$ plane, released from radius 1 AU and heights $z = \pm 0.01$ AU
successively above and below the mid-plane (to avoid crowding in the
figure). The markers mimic a Pointcarr\'e map, i.e. are positioned at
the points where the particle crosses the $\theta = 0$
plane. Particles sizes are respectively: $\blacksquare$ = 0.1mm,
$\blacktriangle$ = 1mm, $\square$ = 1cm, $\vartriangle$ = 10cm,
$\times$ = 1m. {\it Bottom figure}: trajectories of the same particles
in the (z,t) plane, with the same points-style coding for the particle
sizes; again, the points represent positions of intersection with the
$\theta=0$ plane. The analytical trajectories as given by equation
(\ref{eq:epstraj}), using simply the value of $\mu$ at the position
the particle, are shown as dotted lines: they are virtually
indistinguishable from the numerically integrated trajectories. }
\label{fig:epstraj}
\end{figure}

Figure \ref{fig:epstraj} shows the trajectories of small particles
(0.1mm to 1m) released at 1AU from height 0.01 AU, as calculated
numerically using the procedure described in Section
\ref{sec:numint_desc}. It compares the results to those obtained
analytically assuming separation of variables and constant $\mu$ which
are summarized in equation (\ref{eq:epstraj}). The analytical fit is
indistinguishable from the complete numerical solution, despite these
approximations.

\subsubsection{Large particles}
\begin{figure}
\plotone{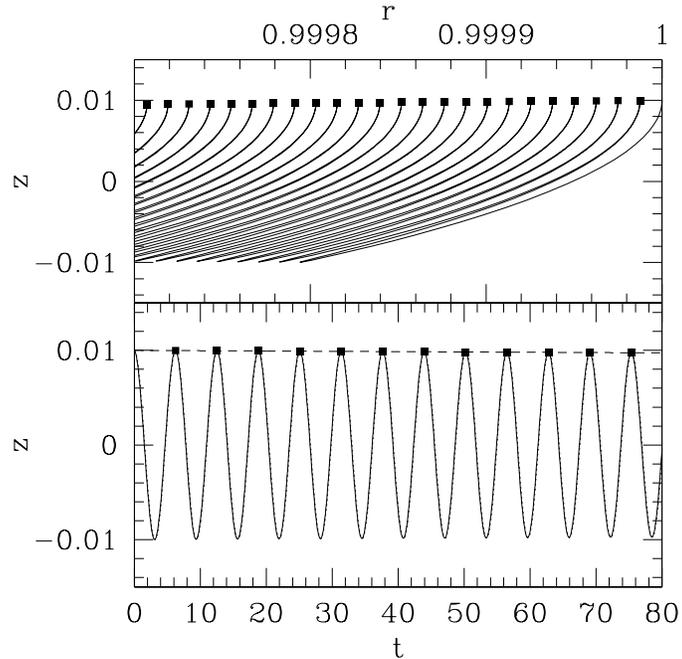}
\caption{{\it Top figure}: trajectory of a 10 m particle in the
$(r,z)$ plane, released from radius 1 AU and height $z = 0.01$ AU
above the mid-plane (to avoid crowding in the figure). The markers
mimic a Pointcarr\'e map, i.e. are positioned at the points where the
particle crosses the $\theta = 0$ plane. {\it Bottom figure}:
trajectory of the same particle in the (z,t) plane; again, the points
represent positions of intersection with the $\theta=0$ plane. The
analytical trajectory as given by equation (\ref{eq:stotraj}), using
simply an average value of $\mu$ in that region ($\mu_{\rm S} = 0.08$)
is shown as dotted lines: they are virtually indistinguishable from
the numerically integrated trajectories. Note that one must take into
account the fact that the normalization of the time-variable varies
with radius in the simple analytical model. The dashed line represents
the envelope of the oscillation as given by equation (\ref{eq:4}).}
\label{fig:stotraj}
\end{figure}

Figure \ref{fig:stotraj} shows the numerically integrated orbit of a
10m sized particle in a protoplanetary nebula, when released from rest
at radius 1AU and height 0.01 above the mid-plane. It compares the
numerical results to the theoretical predictions given by equation
(\ref{eq:stotraj}), and here again, despite the approximations, the
analytical fit is excellent. Note that integration for much longer
times (about 1000 orbits) reveals a small but growing discrepancy in
the amplitude of the oscillation. This is due to the
mis-representation of the analytical expression for the amplitude of
the drag force compared to its true value (see equations
(\ref{eq:realF}) and (\ref{eq:badF})). This discrepancy is discussed
in more detail later.

\subsection{Numerical integration: Boltzmann averaging and collective behavior}

In this section, we now follow the evolution of a collection of 10,000
uniformly sized particles, in the Epstein regime (of size 1mm) and in
the Stokes regime (with size 10m). We release all particles from a
small interval in radius, for $0.9 < r< 1.1$. The initial radial
positions are uniformly distributed in that interval. Other initial
conditions depend on the regime studied, in order to be best able to
perform the adequate comparisons between theory and numerical
experiment. After a time $t$, the particles positions and velocities
are gathered, and binned into regular intervals in radius and in
height. The total number of particles, the average velocity and the
average second velocity moment are then calculated for each bin
according to the formulae:
\begin{eqnarray}
\rho_{\rm p}(r,z) &=& m_p N(r,z) \mbox{ , } \nonumber \\
\overline{w}(r,z) &=& \sum_p w_p / N(r,z) \mbox{ , } \nonumber \\
\overline{ww}(r,z) &=& \sum_p w_p^2 / N(r,z) \mbox{ , }
\end{eqnarray}
where the sums are carried out over all particles found within the
individual bin centered on the position $(r,z)$.

When comparing the numerical data to the analytical solutions of the
previous sections, we are careful to apply the correct normalization
to the analytical formulae: indeed the normalization of the time
variable depends on radial position in the analytical solutions,
whereas in the numerical computations, it is normalized to the orbital
frequency at 1AU.

\subsubsection{Epstein regime}

\begin{figure}
\plotone{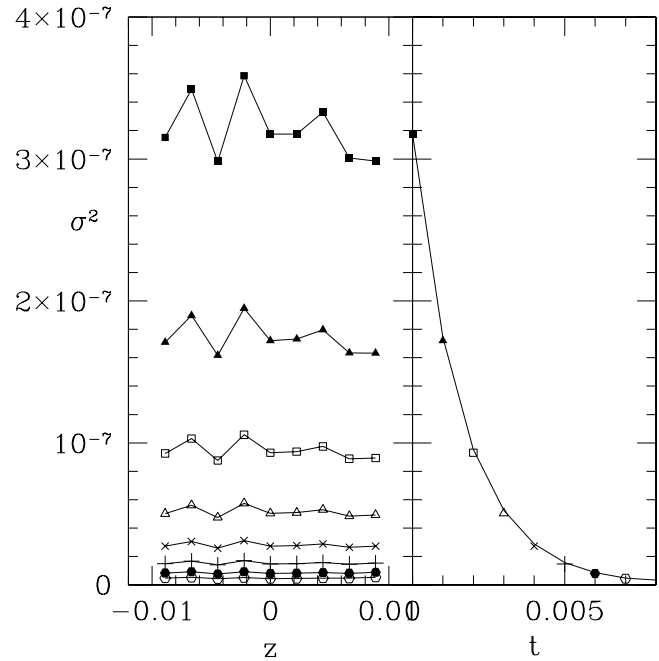}
\caption{{\it Left figure:} Evolution of the velocity dispersion
profile for a collection of mm-size particles. The style of the points
used corresponds to those in the right-side figure. {\it Right figure}
Evolution of the velocity dispersion of the particles located around
$r=1$. The points correspond to the results of the numerical
experiment, and the solid line is the analytical prediction of
equation (\ref{eq:epscomp}), using a value of $\mu$ which is an
average value for this quantity around $r=1$.}
\label{fig:stateps1}
\end{figure}

Using the normalizations adequate to the numerical calculations, the
theoretical solutions for the average particle velocity and dispersion
becomes
\begin{eqnarray}
\overline{w} &=& -\frac{z r^{-3/2}}{\mu(r)}  + O(\exp(-\mu(r) r^{3/2}
t) \mbox{  ,  } \nonumber  \\ \sigma^2 &\propto&  O(\exp(- 2 \mu(r) r^{3/2}  t) \mbox{  .  }
\label{eq:epscomp}
\end{eqnarray}
Figure \ref{fig:stateps1} shows the very short time evolution of a
collection of 10,000 mm-size particles released with a uniform
distribution in heights in the interval $[-0.01,0.01]$ AU and a
uniform distribution in velocities with $ -1.d-3 <w_i < 1.d-3$ (in
units of the Keplerian velocity at $r=1$). For this figure, the
binning in the averaging process is fairly coarse (10 bins in each
direction) in order to generate adequate statistics and to visualize
the results more clearly. The velocity dispersion is indeed found to
decay exponentially in accordance with equation (\ref{eq:epscomp}).

\begin{figure}
\plotone{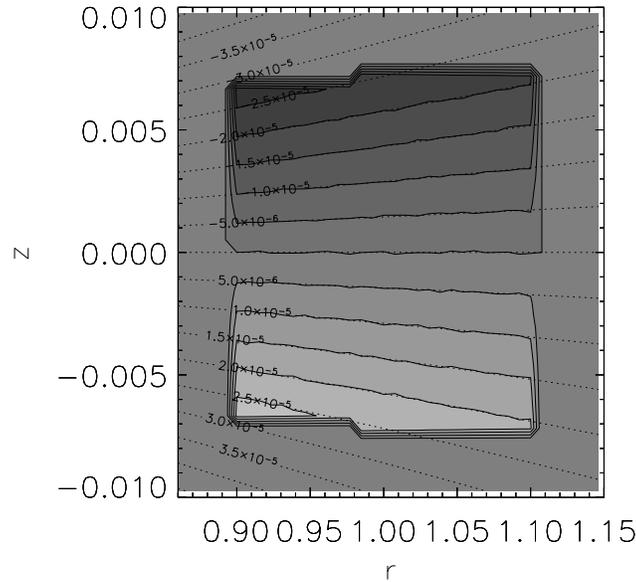}
\caption{Contour plot of the average particle velocity for a
collection of 10,000 mm-size particles. The solid contours as well as
the gray-scale correspond to the numerical results, and the dotted
lines correspond to the analytical predictions of (\ref{eq:epscomp}),
using the local value of $\mu(r,z)$ for each bin. Particles were
released with a uniform distribution of heights in the interval
[-0.01,0.01] AU and of radii in the interval [0.9,1.1] AU. One can
easily observed the settling and (negligible) radial drift in this
diagram, as well as the excellent matching between theory and this
numerical experiment.}
\label{fig:stateps2}
\end{figure}

The trajectory of these particles are then followed up to time $t=100$
(i.e. about 16 orbits) and their average velocity is computed with a
finer binning (50 bins in each direction). The results are presented
in Figure \ref{fig:stateps2}, and compared to the analytical solutions
in equation (\ref{eq:epscomp}). Once again in this regime, the
analytical solutions are found to match the numerical results very
precisely.

\subsubsection{Asymptotic Stokes regime}

In the asymptotic Stokes regime, particles oscillate across the
mid-plane with an algebraically decaying amplitude. Theoretical
analysis of the particles' trajectories and velocities shows 
the velocity dispersion should remain important
at all times.

\begin{figure}
\plotone{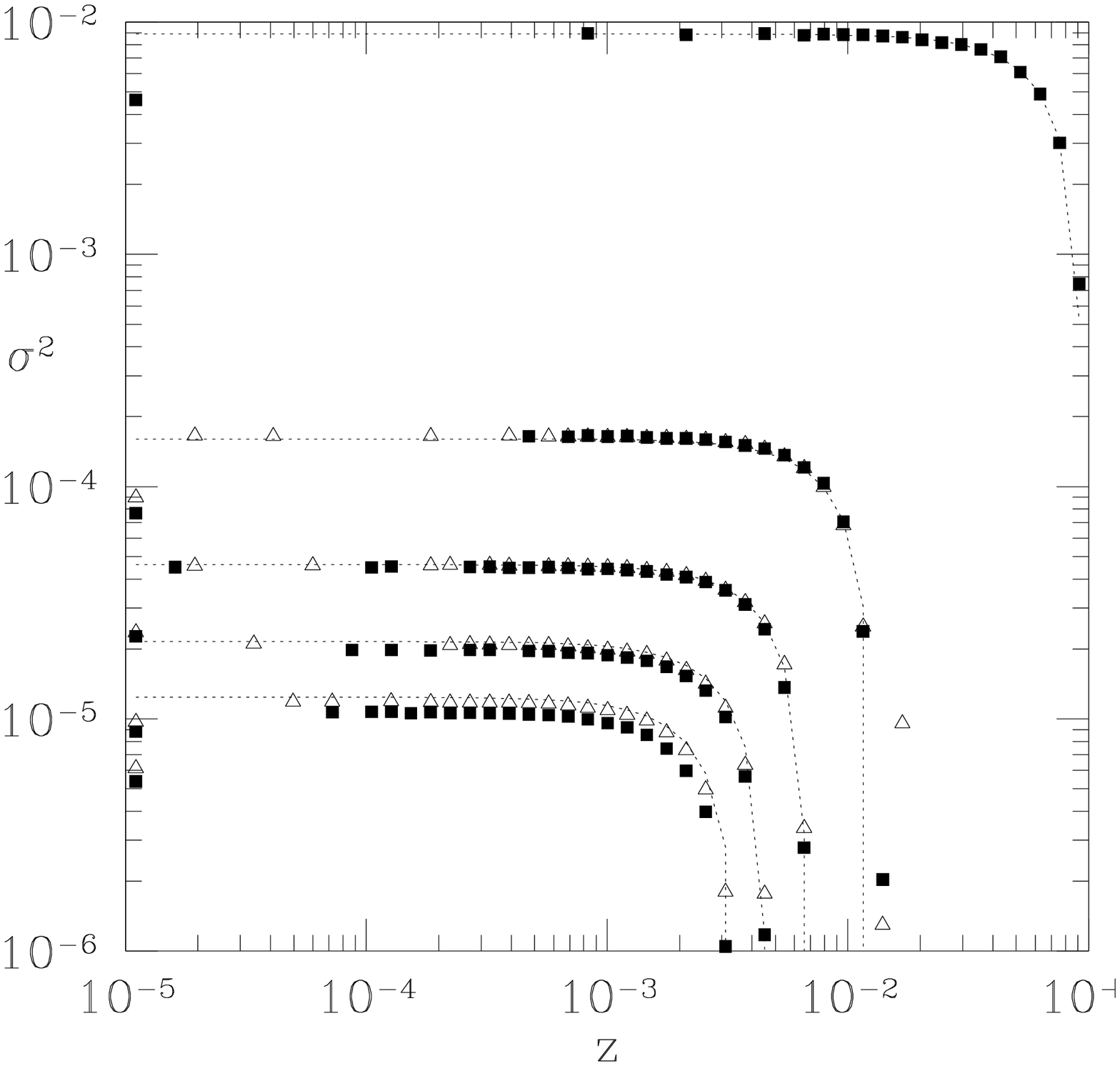}
\caption{Temporal evolution of the velocity dispersion of a collection
of 10m-size particles, measured at $r=1$ AU . The successive curves
(from top to bottom) correspond to times $t=0$, $t=2000$, $t=4000$,
$t=6000$ and $t=8000$ in units of the angular period at this
radius. The following style coding has been used: ($\blacksquare$) for
the realistic numerical experiment and ($\vartriangle$) for the
artificially reduced drag force (as given by equation
(\ref{eq:fakeF})). The dotted lines correspond to the analytical
predictions of equation (\ref{eq:stopred}).}
\label{fig:statsto}
\end{figure}

In order to simplify the comparison between the analytical solutions
in Section \ref{sec:sto} and the full numerical results, we begin by
releasing 10,000 10m-size particles in the disk with similar initial
conditions as the ones that were used in the analytical work. Though
these initial conditions may seem perhaps artificial, they are the
only ones that actually provided any analytical solution. We specify
the initial values
\begin{eqnarray}
z_i &=& \frac{3\pi}{K} \cos(\varphi_i)\mbox{ , } \nonumber \\ w_i &=&
-4\mu_0 \frac{3\pi}{K^2} \cos(\varphi_i) - \frac{3\pi}{K}
\sin(\varphi_i) \mbox{ , }
\label{eq:initzw}
\end{eqnarray}
where $\varphi_i$ is uniformly distributed in the interval
$[0,2\pi]$. The azimuthal velocity of the particles is the Keplerian
velocity at the point of release. We choose these initial values in
order to simulate a set of particles released with a large range of
velocities and heights above the mid-plane.  The gravitationally bound
particles converge toward the mid-plane during the settling phase, and
continue to oscillate coherently across the mid-plane with a random
phase but roughly all the same slowly decaying amplitude. The
simulation is then started at $t=0$ when all the particles have an
oscillation amplitude $3\pi/K$, the corresponding velocity as given by
equation (\ref{eq:initzw}) with a random phase. In that case, using
the same normalization we adopted in our the numerical simulations, we
find the theoretical solutions for the velocity dispersion to be
\begin{equation}
\sigma^2 = \left(\frac{9\pi^2 }{(4\mu(r)t+K)^2} - z^2\right)r^{-3}\mbox{  .  }
\label{eq:stopred}
\end{equation}
Figure \ref{fig:statsto} compares the results of the numerical
calculations to the theoretical solutions given by equation
(\ref{eq:stopred}). For relatively short time after the onset of the
calculation, the numerically obtained dispersion is indeed extremely
well approximated by the analytical formula. However, over much longer
times, a systematic shift emerges, in which the numerical values are
slightly lower than the theoretical solutions. This slight
overestimate of the velocity dispersion comes from the fact that the
analytical solutions underestimate the total drag force by neglecting
the contributions from the particles' motion in the radial and
azimuthal direction. In order to verify this conjecture, another
 numerical experiment is carried out in which the drag force
is artificially set to
\begin{equation} 
\Fd = m_{\rm p} \frac{\rho(r,z)}{\rhos} C(Re)\frac{R}{s} |\dot{z}|
(\dot{r}\er + (r \dot{\theta} - \vg) \etheta + \dot{z} \ez)\mbox{  .  }
\label{eq:fakeF}
\end{equation}
In that case, the contribution of the drag force in the vertical
direction is exactly that used in analytical calculation. As expected,
the analytical expression is now able to reproduce the experimental
results much more accurately.

Having gained some experience with the Boltzmann averaging methods,
and having moreover established the adequacy of the analytical
approximations to the particles' trajectories in the Stokes regime, we
now attempt to use the results of the numerical simulations to find
the evolution of the particles' dispersion in the case where the
initial conditions are chosen randomly (rather than in such a way as
to allow easy analytical calculations). In the follow up numerical
computation, 10,000 10m-size particles are released with a uniform
initial height and vertical velocity distribution chosen in the
intervals $[-0.01,0.01]$ AU and $[-0.01,0.01]v_{\rm K}(R)$. The
azimuthal velocity of the particles is set to be the Keplerian
velocity at the point of release. With these initial conditions, the
initial velocity dispersion should be uniform within the dust layer.

\begin{figure}
\plotone{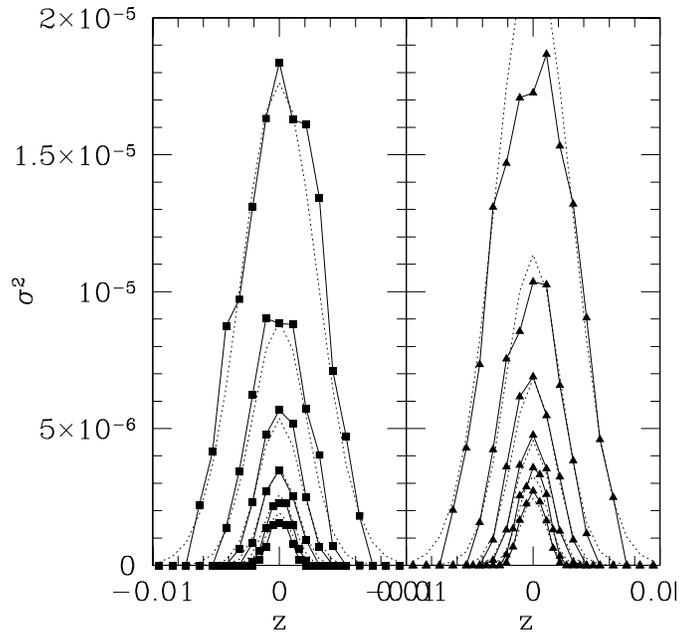}
\caption{{\it Left panel: } Temporal evolution of the dispersion
profile for a set of 10,000 10m-size particles released with a uniform
spatial distribution and a uniform vertical velocity distribution. The
various sets of points joined by a solid line correspond to times
$t=2000$, $t=4000$, $t=6000$, $t=8000$, $t=10000$ and $t=12000$ (from
the top-curve to the bottom curve). The initial velocity dispersion
profile is not shown in order to preserve a readable scale on this
plot.  But it is approximately uniform in the dust layer as expected
. The dotted lines are the corresponding analytical fits to the
dispersion profile as given by equation (\ref{eq:sigmafit1}). {\it
Right panel:} Temporal evolution of the dispersion profile for a set
of 10,000 10m-size particles released with a uniform spatial
distribution and a uniform vertical velocity distribution, with the
artificial drag force amplitude given by equation
(\ref{eq:fakeF}). The curves are plotted at the same time-intervals as
in the left-side panel. The analytical fit, as prescribed by equation
(\ref{eq:sigmafit2}), is shown in dotted lines.  }
\label{fig:storand}
\end{figure}

Figure \ref{fig:storand} shows the evolution with time of this
dispersion profile at radius $r=1$ AU, when the particles are binned
in 20 intervals in radius and heights. After a brief adjustment phase
(not shown in the figure), the dispersion profile 
seems to be well approximated by a Gaussian distribution. We
attempt to match such a Gaussian profile to the numerical data points,
and find that the best fit is given by
\begin{eqnarray}
\Delta(t) &=& \frac{3\pi}{4\mu t + 3\pi/\Delta_0} \mbox{   ,   } \nonumber  \\
\sigma^2(z,t) &=& \frac{\Delta^2(t)}{2}
\exp\left(-\frac{2z^2}{\Delta^2(t)}\right)\mbox{  ,  }
\label{eq:sigmafit1}
\end{eqnarray} 
where $\Delta_0= 0.01$. This formula was obtained from the following
heuristic arguments: according to the formula for the particles'
vertical motion (see equation \ref{eq:stotraj}), the typical vertical
velocity of the particles near the mid-plane is of the order of
$\Delta(t)$, where $\Delta$ is the maximum height of the particles
above the mid-plane. Moreover, $\Delta$ decays with time due to gas
drag according to equation (\ref{eq:4}), so that one can readily
deduce the evolution of $\Delta(t)$ for a set of particles released at
$t=0$ from typical heights ranging between 0 and $\Delta_0$. The
numerical factors have been fitted to the data. The analytical fit is
also shown in Figure \ref{fig:storand}, and is found to reproduce the
evolution of $\sigma$ satisfactorily.

Finally, recalling that the analytical expression for the particles'
vertical motion slightly underestimates the true drag force, we
compare the analytical fit given by equation (\ref{eq:sigmafit1}) to
an artificial simulation in which the drag force is given by equation
(\ref{eq:fakeF}). In principle, the fit should be much better in this
case.  However, we find that the evolution of the dispersion in that
case is better fitted by
\begin{eqnarray}
\sigma^2(z,t) &=& \frac{2}{\pi} \Delta^2(t)
\exp\left(-\frac{2z^2}{\Delta^2(t)}\right)\mbox{  ,  }
\label{eq:sigmafit2}
\end{eqnarray} 
whilst keeping the expression for $\Delta(t)$ the same (see Figure
\ref{fig:storand}). Though this difference is small, it still suggests
that any attempt to find an analytical fit to the collective motion of
Stokes particles should be carefully calibrated before being used in
further simulations.

To conclude this Section, it was found that the analytical evaluation
of one particle's orbit was extremely useful toward the determination
of the collective motion of a large number of particles, and that the
resulting averaged equations fit the numerical experiments extremely
well. We have also been able to determine a simple heuristic way of
closing the moments equations at low order in the case of particles in
the Stokes regime, which emulates the effect of the particles
dispersion by a mock pressure term that {\it can} be related to
large-scale properties (and initial conditions) of the fluid.

\section{Evolution of the size distribution of particles in a gaseous disk}

Having discussed the time-evolution of a large number of particles of
equal size, we now consider a population of different-size
particles. Again, the Boltzmann approach proves useful and simple to
use. Let us assume for simplicity that for a given size, and at a
given radius, the {\it initial} particles to gas mass ratio is
independent of $z$. Hence the initial distribution of particles of a
given size $s$ at time $t=0$ is
\begin{equation}
\rho_i(s,z) = \rho_0(s) \exp(-z^2/2H^2) \mbox{  ,  }
\label{eq:6}
\end{equation}
where $H$ is the typical gas disk height; for an isothermal disk, $H =
c / r\ok$. Let us also assume (again for simplicity) that the initial
vertical velocity of particles is null. This idealized {\it ad hoc}
setup could correspond to a condensation front where, at time $t=0$,
both hydrogen and heavy elemental gas diffuses into a cool region of
the disk. The condensation and sublimation time scales are generally
much shorter than the dynamical time scales (Supulver \& Lin 2000) so
that, at least the small grains may form with a similar density
distribution as the gas and without any initial in-falling motion.
The coexistence of large and small particles well above the mid plane
requires runaway coagulation of initially steady, microscopic dust
grains into grains up to a size $s$.  This rapid {\it in situ} growth
is possible in regions where the particles have large surface density
and much greater than unity area filling factor because the particles'
collision and growth time scales are short compared with their
dynamical time scale.  The particles then start settling from their
initial positions, and interact with the gas mostly with an Epstein
drag law.

The particles density distribution
at time $t$ is then simply:
\begin{equation}
\rhop(s,z,t) = \frac{1}{\alpha} \rho_i(z/\alpha) = \rho_0(s)
\exp(-z^2/2\Delta^2) \exp(t/\mu(s))\mbox{  ,  }
\end{equation}
where $\alpha$ is defined by equation (\ref{eq:5}), and 
$\Delta = H \alpha = H \exp(-t/\mu(s))$ is the evolving dust 
layer thickness. This
equation was derived from (\ref{eq:3}) using the ansatz (\ref{eq:6})
 and $g(w_i,z_i) = \delta(w_i)$. We can deduce from this result the number 
distribution
of particles of size $s$
\begin{eqnarray}
n(s,z,t) &=& n_0(s) \exp(-z^2 /2\Delta^2) \exp(t/\mu(s))  \nonumber \\
&=& n_0(s) \exp(-z^2 \exp(2ts/\kappa)/2 H^2) \exp(ts/\kappa)  \mbox{  ,  }
\end{eqnarray}
where we have defined the constant $\kappa$ such that $\mu = \kappa/s$
in order to write explicitly this equation in terms of its dependence
on the particle size. The initial particle size distribution function
is taken to be that given by Hellyer (1970) and Mathis {\it et al.}
(1977)
\begin{equation}
n_0(s) \propto s^{-3.5} \mbox{  .  }
\end{equation}

\begin{figure}
\plotone{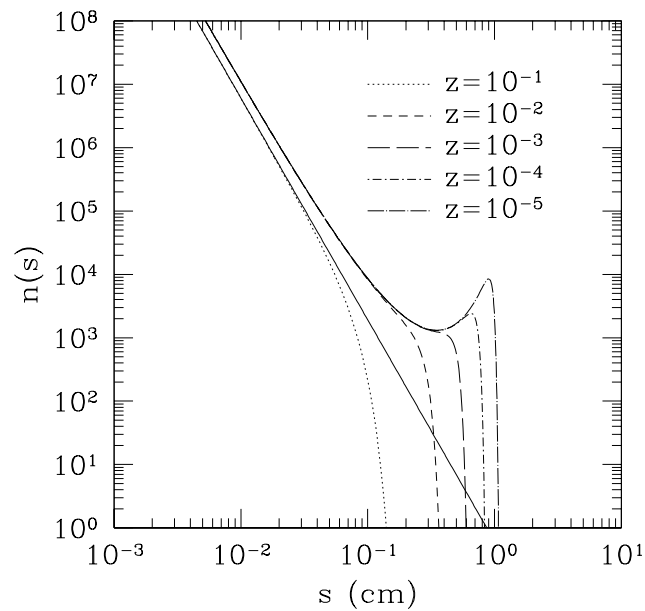}
\caption{Variation with height of the particle size distribution
$n(s)$ at fixed time $t = 10^4$. }
\label{fig:nheight}
\end{figure}

At a given height above the disk, this distribution function peaks for
small particle sizes (which corresponds to the initial peak in
$n_0(s)$), but has another maximum, which evolves with time, at
approximately
\begin{equation}
s = \frac{2\kappa}{t} \ln \left(\frac{H}{z}\right) \mbox{  ,  }
\label{eq:7}
\end{equation}
regardless of the initial size distribution function of the
particles. This result is illustrated in Fig. \ref{fig:nheight}.
\begin{figure}
\plotone{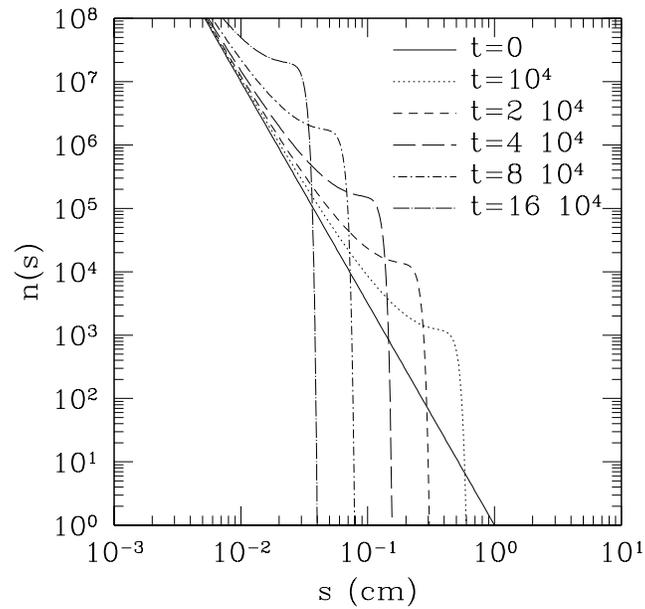}
\caption{Evolution with time of the particle size distribution $n(s)$
at fixed height $z = 0.001$ (i.e. 1\% of the disk scaleheight). Note
how the maximum of the distribution function satisfies the relation
(\ref{eq:7}). }
\label{fig:ntime}
\end{figure}

These results shown in Figures \ref{fig:ntime} and \ref{fig:nheight}
suggest the following comments. Firstly, at a given height, successive
fronts of particles with decreasing sizes pass through the gas as they
settle, to leave the region completely depleted of the intermediate
sized particles. Secondly, at a given time, there is a definite
stratification of the particles in terms of their respective sizes,
with the larger particles concentrating tightly around the
mid-plane. Note that this method is only valid for particles with
sizes up to the order of the mean-free-path of the gas.  For the very
large particles, the Stokes drag law must be applied. In that case, we
have shown in Section \ref{sec:sto} that the large particles only
settle algebraically toward the mid-plane. As a result, although their
number density is small, the large particles remain important at all
times and at all heights in the disk.

This work shows that after a few settling times (which are of the
order of a few thousands of orbits for cm-sized particles), the bulk
of the disk will be depleted of intermediate sized particles, which
have all settled down to the mid-plane. There remains, at all heights,
very small particles that have not had time to settle, and very large
ones which are in a phase of oscillation around the mid-plane.

\section{Summary and Discussions}

In this paper, we provide a set of analytic solutions to describe the
sedimentation of dust particles due to gas drag in a protostellar
accretion disk.

The main effect of the gas drag is to induce a loss of angular
momentum and linear momentum for the particle, which in turn leads to
orbital decay and vertical settling. Numerical integration of the
particles' trajectories in a protoplanetary accretion disk is
straightforward if we assume that the gas flow is laminar and occurs
mostly in the azimuthal direction. However, analytical integration is
also possible given a set of assumptions (which have been checked here
to be reasonably well justified), including the separability of the
radial and vertical motion, and locally uniform density. Accordingly,
we have extracted analytical formulae for the vertical motion of
particles in a disk, both in the Epstein regime (for particles with
size much smaller than the mean-free-path of the gas molecules) and in
the Stokes regime (for particles much larger than the
mean-free-path). Comparison of these analytical expressions with the
complete numerically integrated orbits shows an excellent match in
both cases.

By using the analytical expression for the particles' orbits, we were
able to extract their average behavior using a standard Boltzmann
analysis (i.e. local averaging, or rather, coarse-graining, of the
particles' positions and velocities in phase space). The successive
moments of the Boltzmann distribution function provide the evolution
of the coarse-grained particle density, velocity and velocity
dispersion. The successive moments of the Boltzmann equation yield the
equivalent of ``fluid'' equations for the collection of particles, and
in particular a mass conservation equation and a momentum equation.

Numerical experiments were performed in which 10,000 single-size
particles were released in the disk from a localized region of
phase-space (typically uniformly distributed in space and with a
uniform vertical velocity distribution). Using a similar
coarse-graining method as the one used in the analytical analysis, we
were able to extract the particle density profile, their average
velocity and velocity dispersion from the numerical integration. We
compared them to the analytical formulae obtained previously. The main
results are summarized here.

In the Epstein regime, there is an excellent agreement between the
analytic formulae and the results of the numerical
calculation. Individual particles have their velocity with respect to
the gas quickly damped out on a very short stopping
timescale. Correspondingly, their collective behavior is coherent: the
typical velocity dispersion is found to decay on the particle stopping
timescale, which is usually much smaller than the orbital
timescale. For times longer than the stopping timescale, individual
particles settle exponentially toward the mid-plane and the
corresponding average momentum equation reduces to a simple standard
fluid equation with a linear drag term, and null pressure.

In the Stokes regime, there is a good agreement between the theory and
the results of the numerical experiments. Individual particles are
found to oscillate across the mid-plane with an algebraically decaying
amplitude. Although this behavior is very well reproduced by the
analytical work, there exists a small discrepancy between analytical
formulae and the results of numerical computation in the form of a
slight systematic overestimate of the particles' oscillation
amplitude. This problem is due to an (unfortunately unavoidable)
over-simplification of the amplitude of the drag force in the
analytical work, and results in similar discrepancies in the
comparison of averaged quantities. Nonetheless, apart from this small
identifiable and controllable error, the predictions of the analytical
work match very well the numerical results. As expected, the
particles' velocity dispersion is found to remain large at all times
within the dust layer, with a maximum near the mid-plane. This large
dispersion is directly related to the continuous oscillatory motion of
the particles across the mid-plane. We found it extremely difficult to
obtain analytical predictions of the temporal evolution of the
dispersion for anything but the simplest initial conditions - which
were not necessarily always realistic. However, we were able to deduce
from the numerical experiments a heuristic fit to the observed
evolution of the particles dispersion, for more general initial
conditions. This result has interesting consequences with respect to
the possibility of closing the moments equations at low order, and
describing accordingly the evolution of a collection of large
particles with averaged fluid equations. We will test this theory in
future work.

Finally, by combining the average evolution equations for single-size
particles obtained in the Epstein regime to a plausible
size-distribution function for dust-grains in the ISM (Hellyer, 1970,
Mathis {\it et al.}, 1977), we showed that it is possible to follow
analytically the spatial and temporal evolution of such a
size-distribution as the particles settle toward the mid-plane. This
analytical calculation was done for collisionless particles, and the
extension of this work to include collisions, coagulation and
sublimation would probably require numerical analysis. Nonetheless,
within this approximation we found that the sedimentation process
results in a very strong segregation of particles according to their
sizes, within a timescale of a few hundreds of years only. All
intermediate-size particles quickly converge to the mid-plane, leaving
behind only the very small particles, indirectly supported against the
settling by gas pressure, and the very large particles, which keep
oscillating across the mid-plane with a slowly decaying
amplitude. This segregation process may well suggest that within the
very thin dust layer at the mid-plane, one could approximate the
population of dust particles by a single-size population.

In future work, we shall apply the results obtained here to two essential purposes:
\begin{itemize}
\item the description of the particles as a continuum fluid will be used to study the stability of the dust layer against shearing (Garaud {\it et al.}, in preparation), in order to test whether gravitational instability can indeed take place in the dust layer. More generally, we intend to implement this description into existing 3D HydroDynamics codes for the computation of accretion disks, in order to obtain a fully self-consistent description of dust evolution and growth in these disks;
\item the temporal and spation evolution of the size-distribution of particles can be compared directly with observations of dust. We plan to construct models with which we can infer the extent of particle coagulation and infer the surface density distribution of the gas from high resolution, multi-wavelength (mostly in the sub-mm and mm range) maps of protostellar disks.
\end{itemize}

\section*{Acknoledgements}
Pascale Garaud thanks New Hall (Cambridge) and PPARC for financial support towards the completion of this work. It was also supported in part by NASA through  grant NAG5-10612 and the California Space Institute. The authors thank Neil Balmforth and Taku Takeuchi for many useful discussions.

\section*{Appendix: Derivation of the properties of radial motion of particles in a gaseous disk}

Neglecting the vertical motion of the particles, the equations of
motion in the radial and azimuthal directions in the Epstein regime
are
\begin{eqnarray}
&& \ddot{r} - r \dot{\theta}^2 = -\frac{GM}{r^2} - \frac{c}{s}
\frac{\rho}{\rhos} \dot{r} \mbox{   ,   } \\ && r \ddot{\theta} + 2
\dot{r}\dot{\theta} = - \frac{c}{s} \frac{\rho}{\rhos} \left(
r\dot{\theta} - v_{\rm g}(r)\right) \mbox{   ,   }
\end{eqnarray}
where $v_{\rm g}(r)$ is the slightly sub-Keplerian azimuthal velocity of the
gas
\begin{equation}
v_{\rm g}(r) = \sqrt{\frac{GM}{r}} + \frac{1}{2}
\sqrt{\frac{r^3}{GM}} \frac{1}{\rho}\frac{\ptl p}{\ptl r} = v_{\rm
k}(r) - \frac{\eta(r)}{\ok(r)} \mbox{   ,   }
\end{equation}
which defines $\eta(r) = -(1/2\rho)\ptl p/\ptl r$.  As in Section 2.3,
we use the normalizations $[r] = R = 1$ and $[t] = \ok^{-1}(R)$ to
obtain
\begin{eqnarray}
&& \ddot{r} - r \dot{\theta}^2 = -\frac{1}{r^2} - \mu \dot{r} \mbox{   ,   } \\ && r
\ddot{\theta} + 2 \dot{r}\dot{\theta} = - \mu\left( r\dot{\theta} -
\frac{1}{r^{1/2}} + \frac{\eta(r)}{\ok^2(r)}\right) \mbox{   .   }
\end{eqnarray}

It is not possible to find simple analytical solutions of the full
system. However, we can assume that deviations from Keplerian motions
are small, and that the particle velocity can be written as
\begin{equation}
r = 1 + \epsilon \mbox{ and } \dot{\theta} = 1 + \dot{\phi} \mbox{   .   }
\end{equation}
The linearized system yields (assuming that the background quantities
vary very little with $r$)
\begin{eqnarray}
&& \ddot{\epsilon} + \mu \dot{\epsilon} - 3\epsilon = 2\dot{\phi} \mbox{   ,   } \\
&& \ddot{\phi} + \mu \dot{\phi} = - 2\dot{\epsilon} - \frac{3}{2} \mu
\epsilon - \mu \frac{\eta}{\ok^2} \mbox{   .   }
\end{eqnarray}
Substituting the first into the second yields 
\begin{equation}
\frac{1}{2}\dddot{\epsilon} + \mu \ddot{\epsilon} + \frac{1}{2}
(1+\mu^2) \dot{\epsilon} = -\mu \frac{\eta}{\ok^2} \mbox{   ,   }
\end{equation}
hence the local solution is
\begin{equation}
\dot{\epsilon} = e^{-\mu t} (a\cos 2t + b \sin 2t) -
2\frac{\mu}{1+\mu^2} \frac{\eta}{\ok^2} \mbox{   .   }
\end{equation}
This solution represents the local difference between the particle
velocity and the Keplerian velocity, in units of the Keplerian
velocity. As in the solution found in Section \ref{sec:sto}, the first
term represents a very quick stopping of the particle on timescale
$1/\mu$ and the second term represents a constant inward drift. The
variation of this drift velocity with particle size is given by the
function $\mu/(1+\mu^2)$ which has a maximum for $\mu = 1$. This
result reproduces the results found by Weidenschilling (1977).


\begin{thebibliography}{}

\bibitem{a76}
Adachi I., Hayashi C., Nakazawa K., 1976, Prog. Theor. Phys., 56, 1756
\bibitem{asl}
Adams, F. C., Lada, C. J., Shu, F.  H. 1987, \apj, 312, 788
\bibitem{bh90}
Balbus, S. A, Hawley, J. F., 1990, \apj, 376, 214
\bibitem{beck}
Beckwith, S.V.W. 1999, in {\it The Origin of Stars and Planetary System},
eds C.J. Lada and N.D. Kylafis, Kluwer Academic, 579
\bibitem{b72}
Boltzmann, L., 1872, Wien Berichten, 66, 275
\bibitem{clarke}
Clarke, C. J., Gendrin, A., \& Sotomayor, M. 2001, MNRAS, 328, 485
\bibitem{dac}
D'Alessio, P., Calvet, N., \& Hartmann, L. 2001, \apj, 553, 321
\bibitem{gw73}
Goldreich P., Ward W. R., 1973, \apj, 183, 1051
\bibitem{h00}
Haisch, K. E. Jr., Lada, E. A. \& Lada, C. J. 2001, \apj,  553, 153
\bibitem{h70}
Hellyer B., 1970, MNRAS, 148, 383
\bibitem{hayashi}
Hayashi, C. Nagazawa, K., Nagagawa, Y. 1985, in {\it Protostars and planets 
II}, eds D. Black and M. Mathews, U Arizona Press, 1100
\bibitem{kl01}
Klahr, H. H., Lin, D. N. C., 2001, \apj, 554, 1095
\bibitem{la92}
Laplace, P.S. de 1796, {\it Exposition du syst\`eme du monde}, Paris
\bibitem{marcy}
Marcy, G. W., Cochran, W. D., \& Mayor, M. 2000, in Protostars and
Planets IV, ed V. Mannings, A. P. Boss, \& S. S. Russell (Tucson:
Univ. of Arizona Press), 1285
\bibitem{mal77}
Mathis J. S., Rumpl W., Nordsieck K. H., 1977, \apj, 217, 425
\bibitem{pollack}
Pollack, J.B., Hubickyj, O., Bodenheimer, P., Lissauer, J.J., Podolak,
M., \& Greenzweig, Y.
1996, Icarus, 124, 62
\bibitem{prosser}
Prosser, C. F., Stauffer, J. R., Hartmann, L., Soderblom, D. R.,
Jones, B. F., Werner, M. W., McCaughrean, M. J. 1994, \apj, 421, 517
\bibitem{s98}
Sekiya, M., 1998, Icarus, 133, 298
\bibitem{sh03}
Shuping, R. Y., Bally, J., Morris, M. \& Throop, H. 2003, \apj 587, 109
\bibitem{sl00}
Supulver K. D., Lin D. N. C., 2000, Icarus 146, 525
\bibitem{ta01}
Takeuchi, T, Artymowicz, P., 2001, \apj, 557, 990
\bibitem{tl02}
Takeuchi, T., Lin, D. N. C., 2002, \apj, 581, 1344
\bibitem{thi}
Thi, W.F., Pontoppidan, K. M., Van Dishoeck, E.F., Dartois, E. \&
d'Hendecourt, L.  2002, A \& A,  394, 27
\bibitem{tru} 
Throop, H. B., Bally, J., Esposito, L. W., McCaughrean,
M. J. 2001, Science, 292, 1686
\bibitem{w77}
Weidenschilling S. J., 1977, MNRAS, 180, 57
\bibitem{w84}
Weidenschilling, S.J. 1984, Icarus, 60,553
\bibitem{wc93}
Weidenschilling S. J., Cuzzi, J. N., 1993, {\it Protostars 
\& Planets III}, 1031
\bibitem{wyatt}
Wyatt, M. C., Dermott, S. F., Telesco, C. M.,
Fisher, R. S., Grogan, K., Holmes, E. K., Pina, 
R. K. 1999, \apj,  527 918
\bibitem{w72}
Whipple F. L., 1972,{\it From plasma to planet}, ed. A. Elvius, Wiley, London
\bibitem{ys02}
Youdin, A. N, Shu, F. H., 2002, \apj, 580, 494
\bibitem{zuck}
Zuckerman, B., Forveille, T., \& Kastner, J. H. 1995, Nature, 373, 494



\end{thebibliography}
\end{document}